\documentclass[11pt]{article} % use larger type; default would be 10pt

\usepackage{fullpage}
\usepackage{amsmath}
\usepackage{amssymb}
\usepackage{amsfonts}
\usepackage{mathtools}
\usepackage{graphicx,color,psfrag}
\usepackage{epsfig}

\usepackage[usenames,dvipsnames]{pstricks}
\usepackage{epsfig}
\usepackage{pst-grad} % For gradients
\usepackage{pst-plot} % For axes

\usepackage{hyperref}
\usepackage{booktabs} % for much better looking tables
\usepackage{array}    % for better arrays (eg matrices) in maths
\usepackage{paralist} % very flexible & customisable lists (eg. enumerate/itemize, etc.)
\usepackage{verbatim} % adds environment for commenting out blocks of text & for better verbatim
\usepackage{subfig}   % make it possible to include more than one captioned figure/table in a single float
\usepackage{fancyhdr}
\usepackage{graphicx} % This should be set AFTER setting up the page geometry

\title{Development of a Flow Solver with Complex Kinetics on the Graphic Processing Units\footnotetext{Distribution A: Approved for public release; Distribution unlimited}}
\author{Hai P. Le\footnote{Research Assistant (\href{mailto:hai.le@ucla.edu}{hai.le@ucla.edu})}\\
\textit{Department of Mechanical and Aerospace Engineering, UCLA, Los Angeles, CA}\\\\
Jean-Luc Cambier\footnote{Senior Scientist} \\
\textit{Air Force Research Laboratory, Spacecraft Branch, Edwards AFB, CA}\\
}
\date{\today}

\begin{document}
\maketitle
\abstract
The current paper reports on the implementation of a numerical solver on the Graphic Processing Units (GPU) to model reactive gas mixture with detailed chemical kinetics. The solver incorporates high-order finite volume methods for solving the fluid dynamical equations coupled with stiff source terms. The chemical kinetics are solved implicitly via an operator-splitting method. We explored different approaches in implementing a fast kinetics solver on the GPU. The detail of the implementation is discussed in the paper. The solver is tested with two high-order shock capturing schemes: MP5~\cite{suresh:mp5} and ADERWENO~\cite{titarev:ader}. Considering only the fluid dynamics calculation, the speed-up factors obtained are 30 for the MP5 scheme and 55 for ADERWENO scheme. For the fully-coupled solver, the performance gain depended on the size of the reaction mechanism. Two different examples of chemistry were explored. The first mechanism consisted of 9 species and 38 reactions, resulting in a speed-up factor up to 35. The second, larger mechanism consisted of 36 species and 308 reactions, resulting in a speed-up factor of up to 40.    

\section{Introduction}
\begin{figure}
\begin{center}
	\input{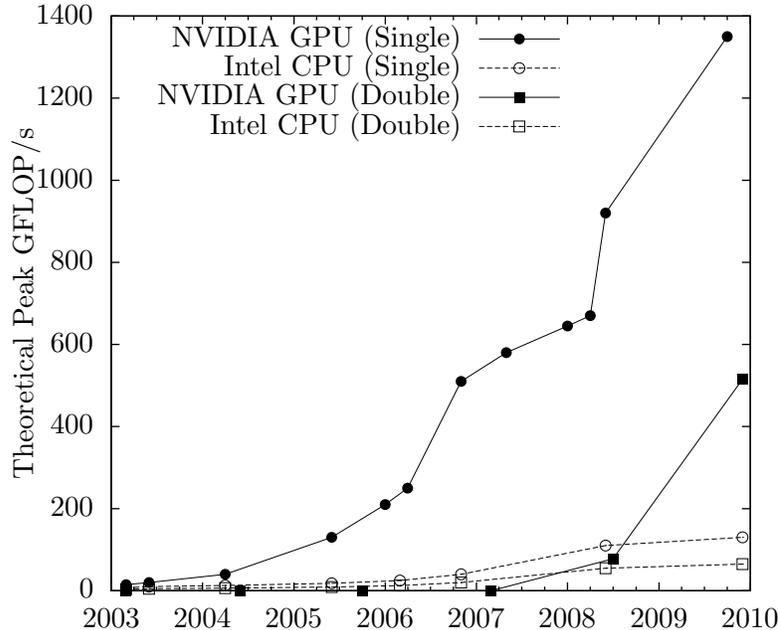}
	\caption{Single and double precision floating point operation capability of GPU and CPU from 2003-2010 (adapted from NVIDIA\cite{CUDA_programming_guide})}
	\label{f:flop}
\end{center}
\end{figure}

During the last seven years, the Graphic Processing Unit (GPU) has been
introduced as a promising alternative to high-cost HPC platforms. Within this period, the
GPU has evolved into a highly capable and low-cost computing solution for scientific
research. Figure~\ref{f:flop} illustrates the superiority of GPU over the traditional Central Processing Unit (CPU)
in terms of floating point calculation. This is due to the fact that
GPU is designed for graphic rendering, which is a highly parallel process. Starting from
2008, the GPU began to support double precision calculation, which is desired for
scientific computing. The newest generation of NVIDIA GPUs called "Fermi" has been
designed to enhance the performance on double precision calculation over the old
generation.
The most popular programming environment for general purpose GPU computing, namely CUDA\cite{CUDA_programming_guide}, has
undergone several development phases and reached a certain level of maturity, which is essential 
for the design of numerical solvers. Several attempts had been made in writing scientific codes on
the GPU, and promising results were obtained both in terms of performance and flexibility~\cite{GPU_Brandvik,GPU_NSSUS, Klockner20097863,Wong20112132}.

In this work, we attempt to follow a similar path on the design of numerical solvers on the GPUs. However, our focus is
different from the previous attempts such that we place more attention to the kinetics solver than the fluid dynamics. 
This is due to the fact that for the simulation of high-speed fluid flow, the computation is dominated by solving the kinetics.
While the current implementation is only for chemical kinetics, we aim to extend it to a more complex and computationally intensive kinetics model for plasma.

\section{Governing Equations}
The set of the Euler equations for a reactive gas mixture can be written as
\begin{equation}
	\label{e:gediv}
	\frac{\partial Q}{\partial t} + \nabla \cdot \bar{F}  = \dot\Omega
\end{equation}
where $Q$ and $F$ are the vectors of conservative variables and fluxes. 
We assumed that there is no species diffusion and the gas is thermally equilibrium (i.e., all species have the same velocity and all the internal energy modes are at equilibrium).
The right hand side (RHS) of equation (\ref{e:gediv}) denotes the vector of source terms $\dot\Omega$, which are composed here of exchange terms due to chemical reactions. 
We solve the system (\ref{e:gediv}) in a finite-volume formulation, by applying Gauss's law to the divergence of the fluxes:
\begin{equation}
	\label{e:gefv}
	\frac{\partial Q}{\partial t} + \frac{1}{V} \oint_S F_ndS = \dot\Omega
\end{equation}
where $Q,F_n$ now denote volume-averaged quantities, which can be written as:
\begin{equation*}
Q = \begin{pmatrix}
			\rho_s \\
			\vdots \\
			\rho u_x \\
			\rho u_y \\
			\rho u_z \\
			E \\
	\end{pmatrix}; 
F_n = \begin{pmatrix}
		\rho_s u_n \\
		\vdots \\
		Pn_x + \rho u_n u_x \\
		Pn_y + \rho u_n u_y\\
		Pn_z + \rho u_n u_z\\
		u_n(P+E) \\
\end{pmatrix}
\end{equation*}
where $u_n$ is the velocity vector normal to the interface and $(n_x,n_y,n_z)$ is the corresponding unit vector.
The total energy is the sum of the internal energies from each species and the total kinetic energy:
\begin{equation}\label{e:energy}
	E = \sum_s \rho_s e_{is} + \frac{1}{2}\rho \vec{u}^2
\end{equation}
Since the species formation energies are not included in that definition, we must account for their change in the source term $\dot\Omega$. 
The hyperbolic terms and the source terms are solved independently of each other by making use of 
an operator-splitting technique~\cite{JLC_supersonic}.
\begin{equation}
	\label{e:operator_splitting}
	\frac{\partial Q}{\partial t} = \left( \frac{\partial Q}{\partial t} \right)_{\text{conv}} 
	+ \left( \frac{\partial Q}{\partial t} \right)_{\text{chem}}
 = - \frac{1}{V} \oint_S F_ndS
+ \dot\Omega
\end{equation}
The equation of state (EOS) is that of an ideal gas, i.e. Dalton's law of partial pressures:
\begin{equation}\label{e:dalton}
P = N R T =\sum_s \rho_s (R/M_s) T 
\end{equation}
where $R$ is the Boltzmann constant (in J/mol$\cdot$K) and $M_s$ the species molar mass.
The pressure can also be determined from the conserved variables, $\{\rho_s,\vec{m},E\}$, where $\vec{m}=\rho\vec{u}$, by:
\begin{equation}\label{e:eos}
	P = (\gamma-1) \left( E- \frac{\vec{m}}{2\rho^2}\right)
\end{equation}
This formulation allows us to compute the pressure derivatives with respect to the conservative variables, needed for the flux Jacobian.
Comparing (\ref{e:dalton}) and (\ref{e:eos}), we find the expression for the \emph{effective} ratio of specific heats $\gamma$:
\begin{equation}\label{e:gamma}
\gamma = 1 + R\,T\frac{\sum_s\rho_s/M_s}{\rho\bar{e}_i}\qquad\text{with   }\rho\bar{e}_i = \sum_s\rho_se_{is}
\end{equation}
Using
\begin{equation}\label{e:deriv_gamma}
\left.\frac{\partial\gamma}{\partial\rho_s}\right)_{E,m}=\frac{RT}{\rho \bar{e}_i}\left(\frac{1}{M_s}-\frac{e_{is}}{\bar{M}\bar{e}_i}\right)
\end{equation}
with $\bar{M}=\rho/N$, we find (using the notation $P_{q_a}=\partial P/\partial q_a$):
\begin{equation}\label{e:deriv_p_rhos}
	P_{\rho_s} = (\gamma-1)\frac{\vec{u}^2}{2} +\left( \frac{RT}{M_s}-\frac{RT}{\bar{M}}\frac{e_{is}}{\bar{e}_i}\right)
\end{equation}
Note that $\sum_s\rho_s P_{\rho_s}\equiv (\gamma-1)\vec{u}^2/2$. The other derivatives are:
\begin{equation}\label{e:deriv_P_m}
	P_{m_\alpha} = -(\gamma-1)u_\alpha \qquad \alpha={x,y,z}
\end{equation}
and
\begin{equation}\label{e:deriv_P_E}
	P_{E} = \gamma-1
\end{equation}
The speed of sound is defined as
\begin{equation}\label{e:sound}
	c^2 = \sum_s \hat{c}_s P_{\rho_s} + (h-\vec{u}^2) P_E = \gamma \frac{P}{\rho}
\end{equation}
where $\hat{c}_s=\rho_s/\rho$ is the species mass fraction and
\begin{equation}\label{e:enthalpy}
	h = \frac{H}{\rho} = \frac{E+P}{\rho}
\end{equation}
is the specific enthalpy.

\section{Numerical Formulation}
\subsection{Fluid Dynamics}
\begin{figure}
\begin{center}
	\scalebox{1} % Change this value to rescale the drawing.
{
\begin{pspicture}(0,-1.0475)(9.0,1.0275)
\psframe[linewidth=0.04,dimen=outer](9.0,1.0075)(0.0,-0.4925)
\psline[linewidth=0.04cm](1.5,-0.4925)(1.5,1.0075)
\psline[linewidth=0.04cm](3.0,-0.4925)(3.0,1.0075)
\psline[linewidth=0.04cm](4.5,-0.4925)(4.5,1.0075)
\psline[linewidth=0.04cm](6.0,-0.4925)(6.0,1.0075)
\psline[linewidth=0.04cm](7.5,-0.4925)(7.5,1.0075)
\psline[linewidth=0.04cm](0.0,-0.0525)(1.5,0.3675)
\psline[linewidth=0.04cm](1.5,0.3675)(3.0,0.5875)
\psline[linewidth=0.04cm](3.0,0.5875)(4.5,0.3875)
\psline[linewidth=0.04cm](4.5,-0.1125)(6.0,-0.0325)
\psline[linewidth=0.04cm](6.0,-0.0325)(7.5,0.1875)
\psline[linewidth=0.04cm](7.5,0.2075)(8.98,0.8675)

\rput(0.7732813,-0.8225){$j-2$}
\rput(2.2696874,-0.8225){$j-1$}
\rput(3.8710938,-0.8225){$j$}
\rput(5.26625,-0.8225){$j+1$}
\rput(6.771094,-0.8225){$j+2$}
\rput(8.269688,-0.8225){$j+3$}
\rput(4.9540626,0.2875){$u^R_{j+\frac{1}{2}}$}
\rput(4.054531,0.1075){$u^L_{j+\frac{1}{2}}$}
\end{pspicture} 
}
	\caption{Schematic of computational stencil with left and right states.}
	\label{f:stencil}
\end{center}
\end{figure}
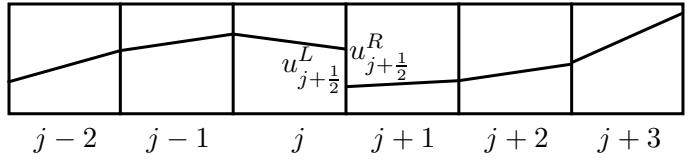

A dimensional splitting technique~\cite{toro:Riemann_solver} is utilized for solving the convective part of the governing equations.
In order to achieve high-order both in space and time, we employed a fifth-order Monotonicity-Preserving scheme\cite{suresh:mp5} (MP5) for 
the reconstruction, and a third-order Runge-Kutta (RK3) for time integration. 
For the MP5 scheme, the reconstructed value of the left and right states of interface $j+\frac{1}{2}$ is given as (see Fig. \ref{f:stencil}): 
\begin{subequations}\label{e:mp5}
\begin{align}
	u^L_{j+\frac{1}{2}} &= \frac{1}{60}\left( 2u_{j-2} -13u_{j-1} + 47u_j + 27u_{j+1} - 3u_{j+2} \right)\\
	u^R_{j+\frac{1}{2}} &= \frac{1}{60}\left( 2u_{j+3} -13u_{j+2} + 47u_{j+1} + 27u_{j} - 3u_{j-1} \right)
\end{align}
\end{subequations}

The reconstructed values are then limited to avoid instability.
\begin{equation}\label{e:mp}
	u^L_{j+\frac{1}{2}} \leftarrow \text{median} \left( u^L_{j+\frac{1}{2}},u_{j},u_\text{MP} \right)
\end{equation}
where
\begin{equation}\label{e:ump}
	u_\text{MP} = u_j + \text{minmod} \left[u_{j+1}-u_{j}, \alpha \left( u_{j}-u_{j-1} \right)\right]
\end{equation}
with $\alpha = 2$.

In addition to the MP5 scheme, we also considered the Arbitrary Derivative Riemann Solver using the Weighted Essentially Non-Oscillatory reconstruction procedure, so-called ADERWENO scheme~\cite{titarev:ader}. At each interface of 1 one-dimensional stencil, we seek the solution 
of the generalized Riemann problem (GRP)
\begin{equation}\label{e:grp}
	\partial_t Q + \partial_x F(Q) = 0
\end{equation}
with the following initial conditions
\begin{equation*}
Q^{(k)}(x,0) = \left\{ \begin{array}{rcl}
q^{(k)}_L(x) & \mbox{if} & x<0 \\ q^{(k)}_R(x)  & \mbox{if}  &x>0
\end{array} \right.
\end{equation*}
The solution of equation (\ref{e:grp}) can be expanded using the Taylor Series expansion in time.
\begin{equation}
	Q(x_{j+1/2},t+h) = Q(x_{j+1/2},t) + \sum_{k=1}^{r-1} \frac{h^k}{k!} \frac{\partial^k}{\partial t^k} Q(x_{j+1/2},t)
\end{equation}
where all the temporal derivatives can be determined using the Cauchy-Kowalewski procedures~\cite{titarev:ader}. 
The solution obtained in this form is high-order both in space and time, so a one-step time integration approach such as the Euler
 explicit method is adequate. This provides certain advantages over the MP5 scheme since the overhead due to RK integration
can be avoided. One disadvantage however of the ADERWENO scheme is that the scheme is not guaranteed to be total variation diminishing (TVD) which might be an issue in the region of strong compression or expansion waves.

The interface fluxes are solved by employing the HLLE Riemann solver~\cite{Einfeldt:HLLE}, which is given as
\begin{equation}
	\label{e:fluxhlle}
	F^{\text{HLLE}}_{j+1/2} = \frac{b^+F_R-b^-F_L}{b^+-b^-} + \frac{b^+b^-}{b^+-b^-} \Delta Q_{j+1/2}
\end{equation}
where
\begin{equation}	
	b^+ = \text{max}(0,\hat{u}_n+\hat{c},u_{nR}+c_R)
\end{equation}
\begin{equation}	
	b^- = \text{min}(0,\hat{u}_n-\hat{c},u_{nL}-c_L)
\end{equation}

\subsection{Chemical Kinetics}
An elementary chemical chreaction takes the form
\begin{equation}
	\sum_s \nu_r' [X_s] \Leftrightarrow \sum_s \nu_r'' [X_s]
\end{equation}
where $\nu_r'$ and $\nu_r''$ are the molar stoichiometric coefficients of the reactants and products of
each reaction. The forward rate can be determined from 
\begin{equation}
	K_{fr} = A_{fr}T^{\beta_r} \exp\left( -\frac{E_r}{RT}\right)
\end{equation}
The backward reaction rate is calculated from the equilibrium constant, which is given as
\begin{equation}
	K_{e} = \frac{K_{fr}}{K_{br}} = \left( \frac{P_a}{RT} \right) ^{\sum_s \nu_s} \exp\left( \frac{- \Delta G^0}{RT} \right)
\end{equation}
For each reaction, the progression rate can be written as
\begin{equation}
	Q_r = \sum_s \alpha_{rs} [X_s] \left[ K_{fr} \prod_r [X_s]^{\nu'_{rs}} - K_{br} \prod_r [X_s]^{\nu''_{rs}} \right]
\end{equation}
The species net production rate can then be determined from
\begin{equation}
	\dot\omega_s = \sum_r M_s \nu_{rs} Q_{r}
\end{equation}
By conservation of mass, sum of all the species production rates should be equal to zero
which yields the following expression.
\begin{equation}
	\sum_s \dot\omega_s = 0
\end{equation}

In order to solve for the change in the species concentration through production and loss
rate, one needs to know all the changes in the thermodynamics for each reaction as well
as their rates. In practice, the backward rate can also be computed using curve-fitting
technique with the temperature as an input, but to be more rigorous, it is recomputed
using the equilibrium constant. All these quantities are
read from separated data files which contain all the species information used for the
computation along with the elementary reactions.

The chemical kinetics is solved using a point implicit solver to ensure stability. The
formulation can be obtained by using a Taylor series expansion in time of the RHS
\begin{equation}
	\frac{dQ}{dt} = \dot\Omega + \Delta t \frac{\partial \dot\Omega}{\partial t}
\end{equation}
By applying chain rule to the time derivatives on the RHS, one could obtain
\begin{equation}
	\label{e:kinetics}
	\left( I - \Delta t \frac{\partial \dot\Omega}{\partial Q} \right) \frac{dQ}{dt} = \dot\Omega
\end{equation}
Equation (\ref{e:kinetics}) is now a system of algebraic equations, which can be solved using a variety of methods.
In the current work, a direct Gaussian elimination procedure is carried out in order to solve
for the linear system of the chemical kinetics. It must be pointed out that the Gaussian elimination procedure
scales with $(N_s)^3$ where $N_s$ in this case is the number of species. Solving the system at every cell is clearly a computationally intensive task.

\section{GPU Implementation}
\begin{figure}
\begin{center}
	\includegraphics[height=8cm]{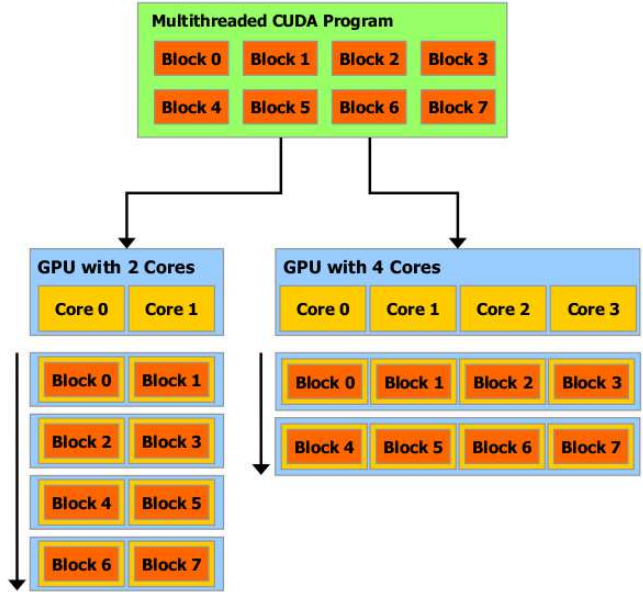}
	\caption{Data parallelism in GPU (adapted from NVIDIA\cite{CUDA_programming_guide})}
	\label{f:multithreaded_CUDA}
\end{center}
\end{figure}

The GPU processes data on Single-Instruction-Multiple-Thread (SIMT) fashion. The instruction for executing
on the GPU is called a \textit{kernel} which is invoked from the host (CPU). The CUDA programming model consists of \textit{grid} and \textit{thread block}. A \textit{grid} consists of multiple \textit{thread blocks} and each \textit{thread block} contains a number of \textit{threads}. When a \textit{kernel} is called, the scheduler unit on the device will automatically assign a group of thread blocks to the number of available streaming multi-processors (SM or GPU core) on the device. Once the SM has completed the calculation, it will be assigned another block. Since there is no communication between the \textit{thread blocks}, the execution order is automatically optimized so GPU with more cores will perform the calculation faster. This is shown in Figure~\ref{f:multithreaded_CUDA}.

The data parallelism is also inherent at the thread level. An instruction given to a \textit{thread block} is handled by a SM which contains a number of streaming processors (SP). All the threads within each block will be organized into groups of 32 threads called \textit{warps} which are executed in a SIMT manner. The difference in the data parallelism between \textit{grid} and \textit{thread block} is that there is synchronization mechanism for all the threads in a same block but not for all the blocks in the grid. It is therefore important to ensure that there is no data dependency between \textit{thread blocks}.

\subsection{Computational Fluid Dynamics}
The parallelization is done by directly mapping the computational domain to the
CUDA grid. The face values can be mapped the same way
with a larger grid since the number of faces in each direction is always 1 greater than the
number of cells in that direction. Each CUDA thread can be associated with one cell/face
inside the computational domain.

In order to maximize the memory access efficiency, the data for the whole domain 
is stored as a one-dimensional array. The index of this array can be 
calculated from the dimensional indices and the variable index and vice versa. This
is similar to the approach taken in \cite{Wong20112132}.  Since all the data storage
is one-dimensional, the domain can be decomposed
into one-dimensional stencils of cells/faces values where each stencil can be assigned to a
CUDA block. Since the computational domain can be up to three-dimensional, one can
split the stencil different ways. However, it is desired to split the stencil so that all the
components of a stencil are located in contiguous memory space. For example, if $i$ is the
fastest varying index of a three-dimensional data array $A(i,j,k)$, the stencil is created by splitting
the domain along the $i$ direction. The directional 
indices $(i,j,k)$ of a three-dimensional domain with lengths $IDIM$, $JDIM$ and $KDIM$ can be calculated from the thread index as follows:

\begin{verbatim}
k  = thread_Idx  /  (IDIM*JDIM)
ij = thread_Idx mod (IDIM*JDIM)
j  =     ij      /   IDIM
i  =     ij     mod  IDIM
\end{verbatim}

Each stencil now can be fitted into a
block of threads and each component of the stencil is associated with a thread. Since all the threads
within a block are accessing consecutive memory address, the access pattern is coalesced
resulting in high memory bandwidth. The calculation inside the \textit{kernel} requires a certain
amount of registers, especially for high-order schemes, so the size of the stencil is only
constrained by the size of the available registers in each warp. However, within that
constraint, the size of the block can have an impact on the performance of the \textit{kernel}.

It is usually recommended to maximize the block occupancy to make up for the memory
latency. Since the occupancy factor is proportional to the block size, a large block size would result in high occupancy. However, as studied by Volkov~\cite{Volkov}, in the case 
where there are multiple independent instructions in the \textit{kernel}, it is more advantageous to 
make the block size smaller and utilize more registers to cover for the memory latency. 
One example in the case of the fluid solver is the construction of
the eigensystem. Since each entry of the eigensystem can be constructed independently of the 
others, the \textit{kernel} actually performs faster in the case of smaller block size. This is referred
as Instruction-level Parallelism (ILP). More information on how to optimize the performance of 
a \textit{kernel} using ILP can be found in Ref. \cite{Volkov}.

\subsection{Chemical Kinetics}
The parallelization of the kinetics solver can greatly benefit from GPU acceleration. The problem 
can be described as a simple linear algebra problem $Ax = b$ where $A$ is the Jacobian matrix mentioned in equation (\ref{e:kinetics}). The solution of the system contains the change in molar concentration of all the species due to chemical reactions. The system is solved using a Gaussian elimination algorithm, which is basically a sequential method. Since the kinetics in each computational cell are independent of other cells, one can parallelize the system on thread-per-cell basis. The Gaussian elimination process requires a lot of memory access for read and write instructions to use and modify the values of the Jacobian. We investigated here different approaches to maximize the performance of the kinetics solver.

The first approach is to store everything on global memory. Although global memory is the slowest type of memory on the device, coalesced memory access can result in high memory bandwidth close to the theoretical limit. The advantage of this approach is that it is simple and straight-forward. The global memory being the largest on the device, the restriction on the number of species can be relaxed. It must be noted that if the number of species is large, the Gaussian method may suffer from accumulated round-off error, and double-precision is rapidly a necessity.

\begin{figure}
\begin{center}
	\input{chem_limit1.tex}
        \includegraphics[scale=1]{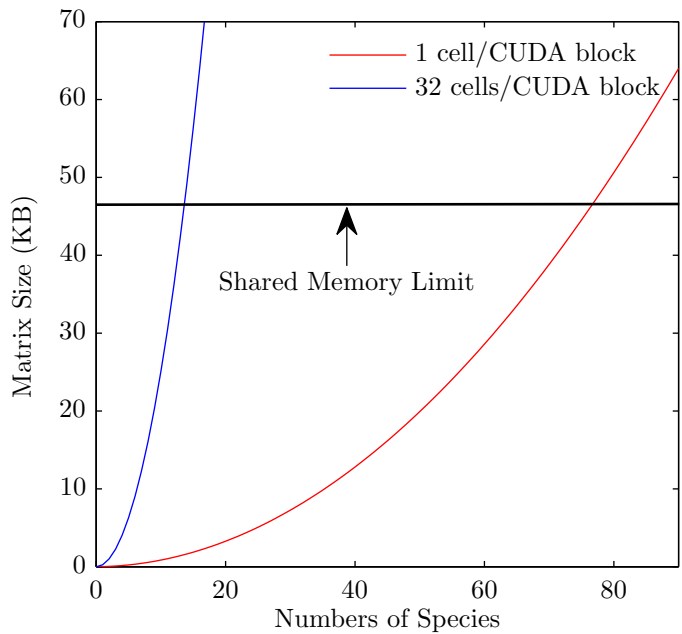}
	\caption{Chemistry size limit}
	\label{f:chem_limit1}
\end{center}
\end{figure}

It is usually recommended~\cite{CUDA_programming_guide} to utilize shared memory whenever possible to reduce global memory traffic. In this case, solving the chemical system requires inverting a $N_s$-by-$N_s$ matrix and the system needs to be solved at every computational cell. Storing the whole Jacobian and the RHS vector on shared memory is not an ideal situation here. Figure \ref{f:chem_limit1} shows the memory requirement for storing the Jacobian and RHS on the typical shared memory (48 KB for a Tesla C2050/2070). If we associated an entire thread block to the chemical system in a cell, the number of species is limited to 75. Storing more than 1 system per block makes this limit go even lower, as can be seen for 1 thread per cell (32 threads block size). 

\begin{figure}
\begin{center}
	\input{chem_limit2.tex}
        \includegraphics[scale=.91]{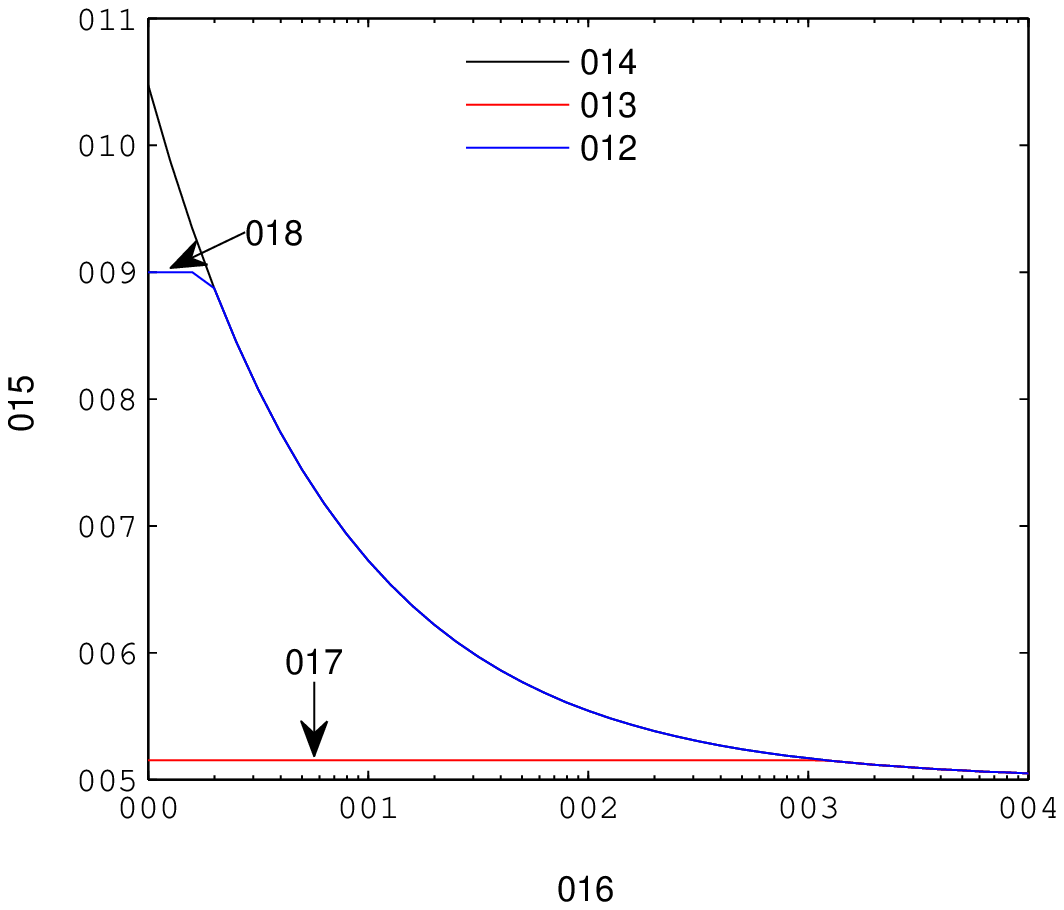}
	\caption{Chemistry size limit}
	\label{f:chem_limit2}
\end{center}
\end{figure}

In order to overcome the shared memory limit, we considered storing only two rows of the Jacobian in shared memory since the sequence of the elimination is done row-by-row. For each row elimination, one need to store values for that row and the pivot row. This is shown in Figure \ref{f:chem_limit2} where the species limit is now much higher than for the full storage pattern. The draw-back of this approach, however, is that there are multiple memory transfers between global and shared memory, since we are required to copy back the values of each row after being eliminated. The parallelization is only effective when the calculation is dominant, so that it would make up for the memory transfer. It will be shown later that this algorithm is only fast for a large number of species.

\section{Results}
\subsection{Solver Results}
\begin{figure}
\begin{center}
	\includegraphics[height=5cm]{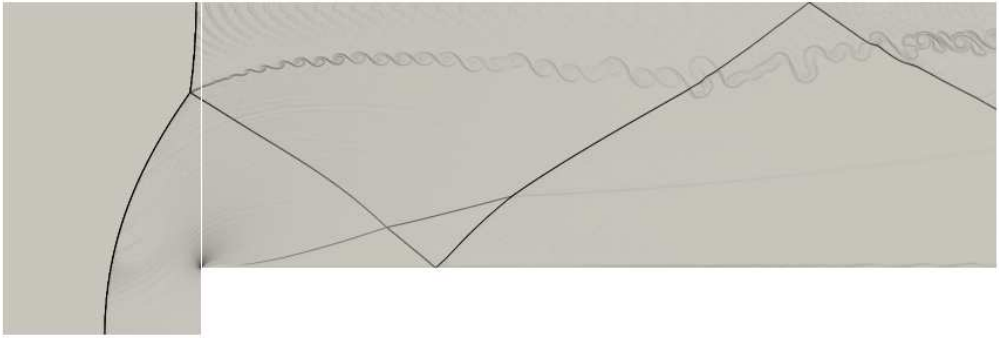}	
	\caption{Solution of the forward step problem}
	\label{f:est}
\end{center}
\end{figure}

The first objective is to verify that the solver is correctly implemented using the CUDA kernels. For this purpose, we can compare the results with a pure-CPU version, but also compute a set of standard test cases.
The first of those is a Mach 3 wind tunnel problem (a.k.a. the forward step problem) using the MP5 scheme, whose solution shown in Figure~\ref{f:est}. This problem had been utilized by Woodward and Colella\cite{Woodward1984115} to test
a variety of numerical schemes. The whole domain is initialized with Mach-3 flow and reflective boundary condition are enforced on the step and the upper part of the domain. The left and the right boundary conditions are set as in-flow and out-flow, respectively. Special attention is usually required at the corner of the step since this is a singular point of the flow which can create numerical instabilities. Woodward and Colella treated this by assuming the flow near the corner is nearly steady. However, this artificial fix was not used in this simulation since we want to test the robustness of the solver in the case of strong shocks and how it handles the singularity in wall curvature, responsible for very strong expansion.

\begin{figure}
\begin{center}
	\includegraphics[height=10cm]{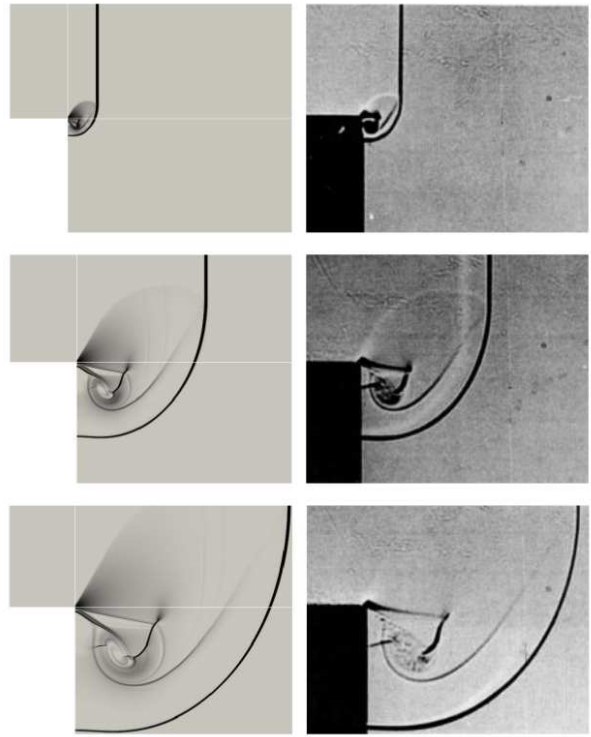}	
	\caption{Diffraction of a Mach 2.4 shock wave down a step. Comparison between numerical schlieren and experimental images}
	\label{f:shockdif}
\end{center}
\end{figure}

The second test involves a similar problem of a diffraction of a shock wave ($M=2.4$) down a step\cite{Vandyke}. The strong rarefaction 
at the corner of the step can cause a problem of negative density when 
performing the reconstruction. The problem is modeled here using 27,000 cells, and the numerical simulation is shown in pair with the experimental images in figure \ref{f:shockdif}. The solver was able to reproduce the correct flow features with excellent accuracy.

We also modeled the Rayleigh-Taylor instability problem \cite{RT}. The problem is described as the acceleration of a heavy fluid into a light fluid driven by gravity. For a rectangular domain of $(0.25 \times 1)$, the initial conditions are given as follows:
\begin{equation*}
	\rho =2, u=0,v=-0.025 \cos (8 \pi x), P=2y+1 \text{ for } 0 \le y \le \frac{1}{2}
\end{equation*}
\begin{equation*}
	\rho =1, u=0,v=-0.025c \cos (8 \pi x), P=y+\frac{3}{2} \text{ for } \frac{1}{2} \le y \le 1
\end{equation*}
where $c$ is the speed of sound.

The top and bottom boundaries are set as reflecting and the left and right boundaries are periodic. As the flow progresses, the shear layer starts to develop and the Kelvin-Helmholtz instabilities become more evident. The gravity effect is taken into account by adding a source term vector which modifies the momentum vector and the energy of the flow based on gravitational force. The source term in this case is relatively simple and contributes very little to the overall computational time. The performance of the fluid dynamics calculation is discussed in the next section of this report.

\begin{figure}
\begin{center}
	\includegraphics[height=15cm]{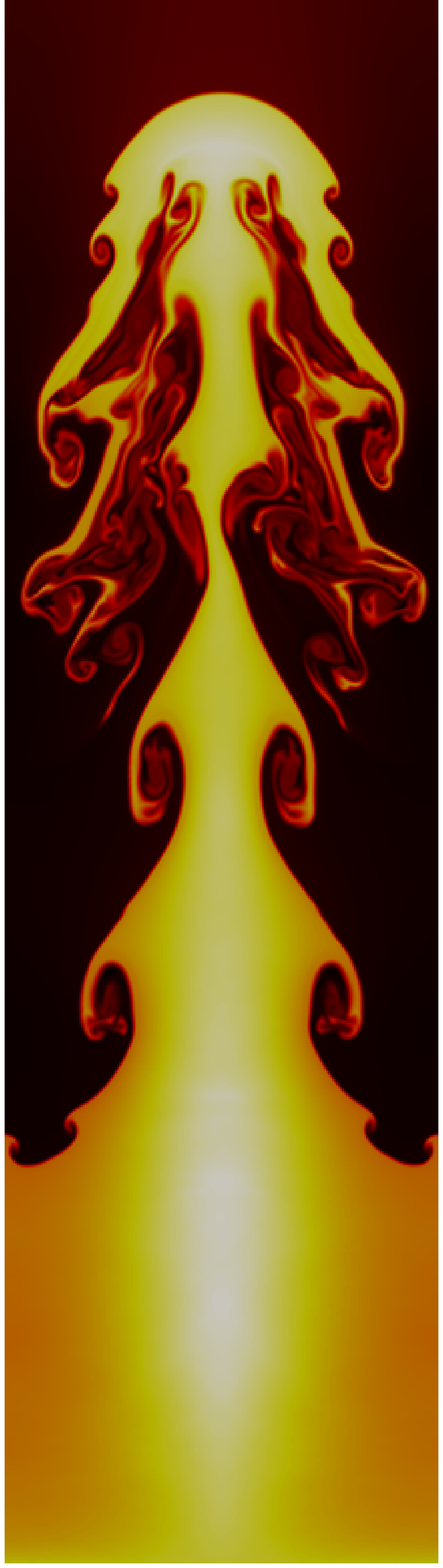}	
	\includegraphics[height=15cm]{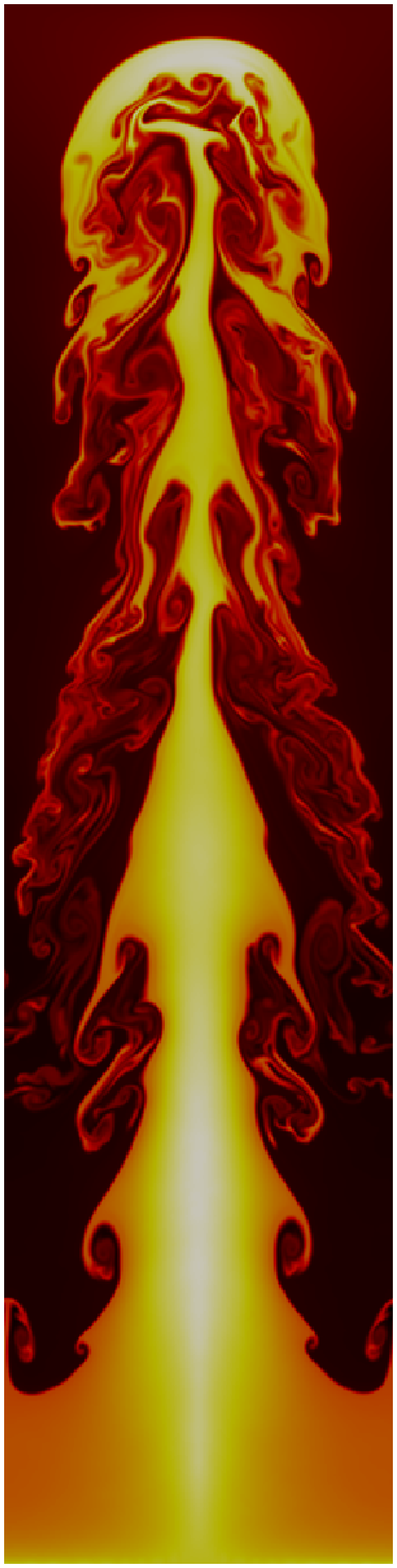}
	\includegraphics[height=15cm]{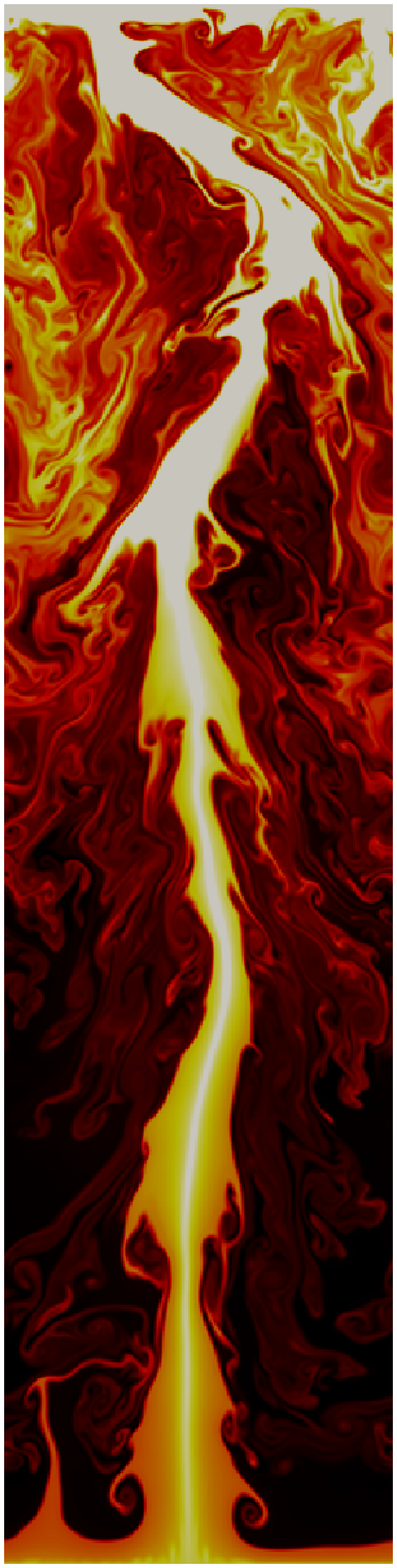}		
	\caption{Rayleigh-Taylor instability computed with the MP5 scheme on a $400 \times 1600$ grid}
	\label{f:rayltay1}
\end{center}
\end{figure}

We now turn the attention to the modeling of a reactive flow field. We simulated a spark-ignited detonation wave both in one- and two-dimension to demonstrate the capability of the solver. At a well-resolved scale, the detonation wave can be described as a strong shock wave supported by the heat release from a high-temperature flame behind an induction zone. Interesting features have been observed both in the 1-D and 2-D simulations, characterized by the coupling of the fluid dynamics and chemical kinetics. The study of flame-shock coupling is an on-going research topic and certainly can be aided with GPU computing when the evolution of the detonation wave needs to be resolved at a very fine spatial scale. 

\begin{figure}
\begin{center}
        \includegraphics[height=2cm]{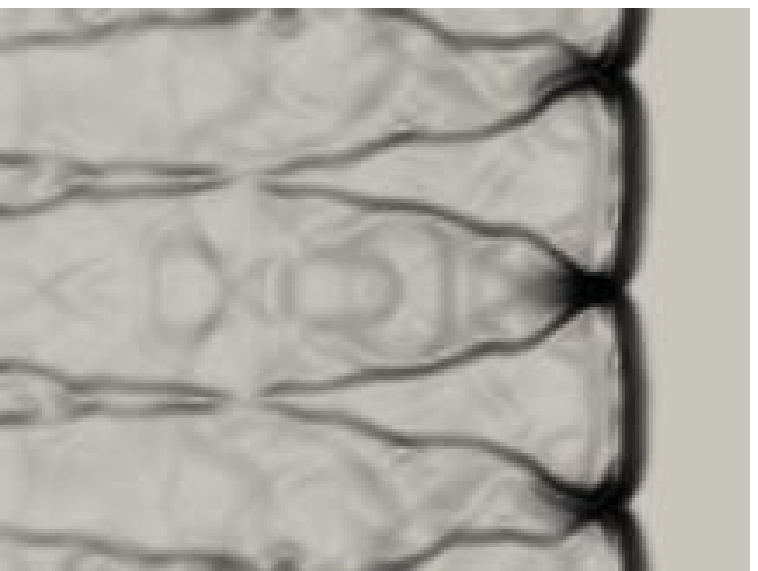}
        \includegraphics[height=2cm]{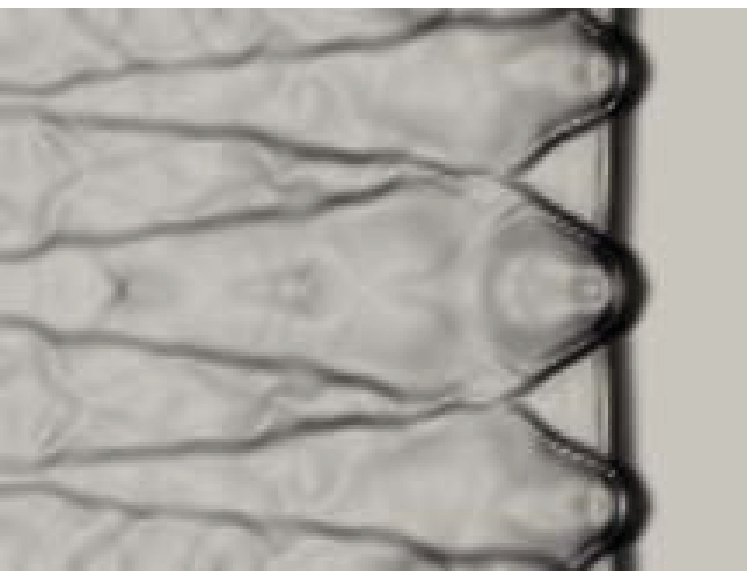}
        \includegraphics[height=2cm]{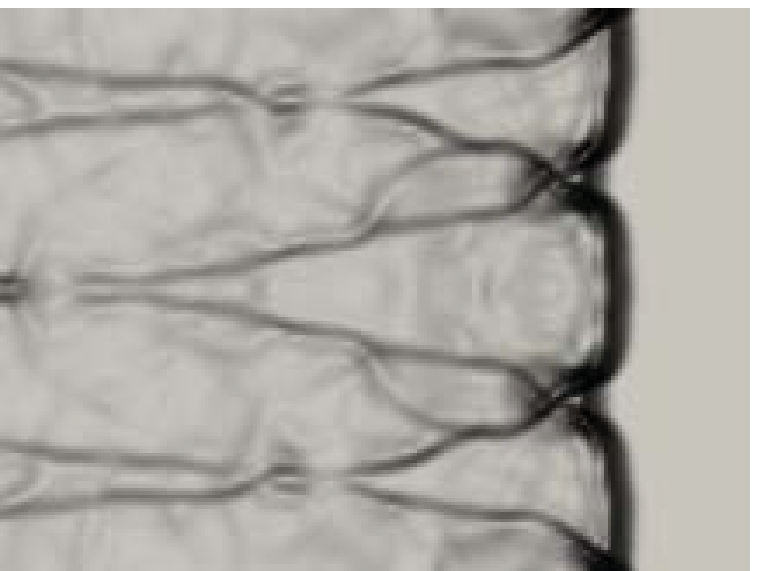}
        \includegraphics[height=2cm]{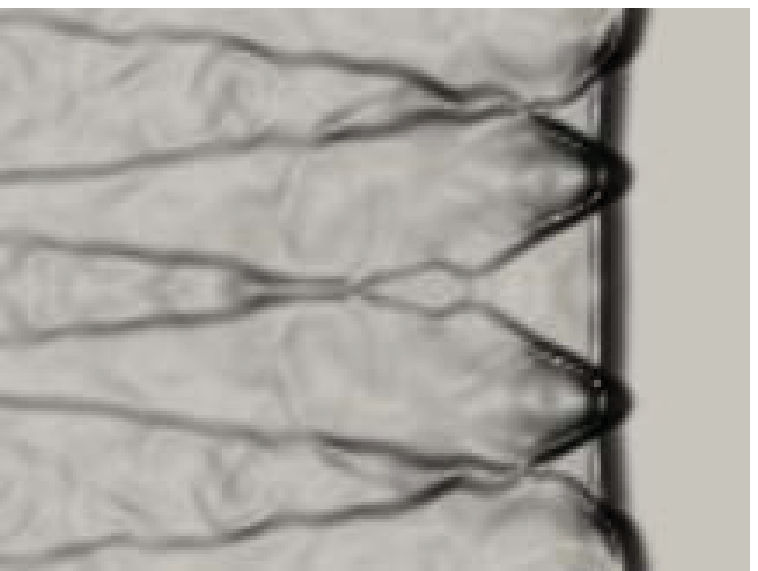}
        \includegraphics[height=2cm]{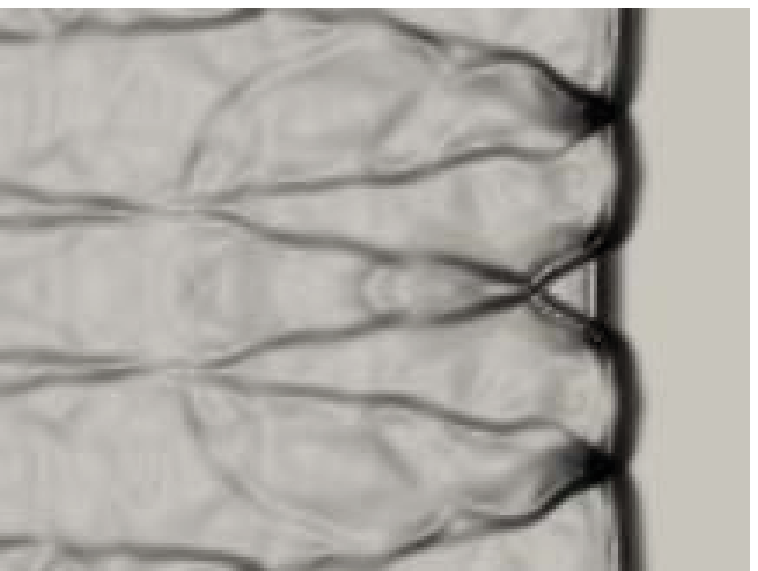}\\
        \includegraphics[height=2cm]{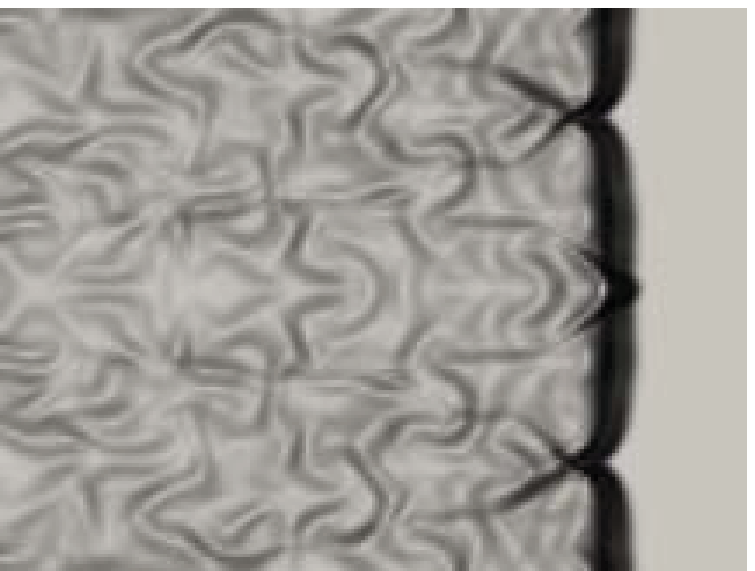}
        \includegraphics[height=2cm]{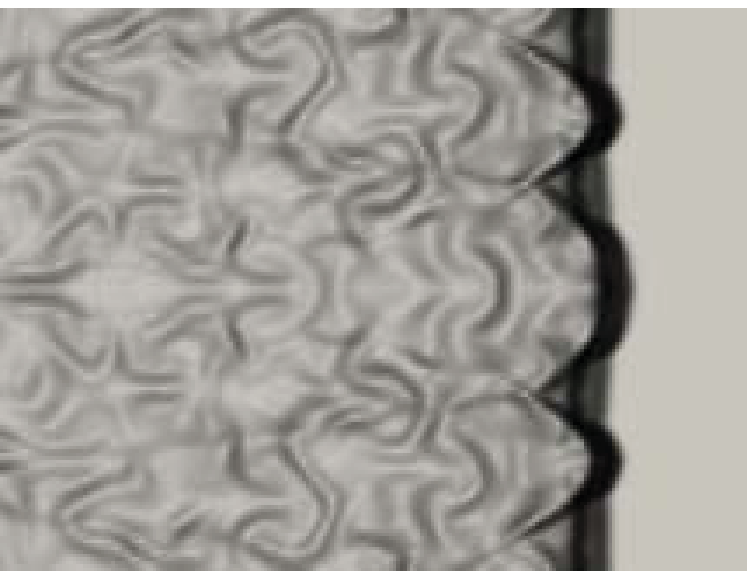}
        \includegraphics[height=2cm]{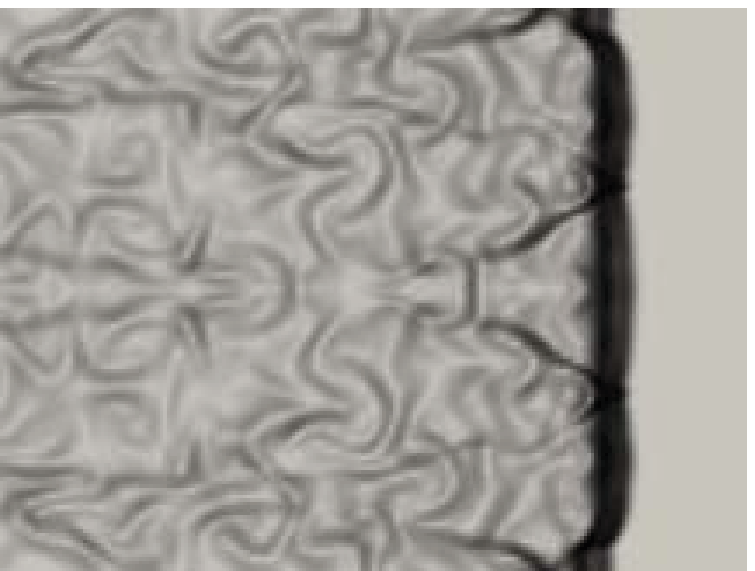}
        \includegraphics[height=2cm]{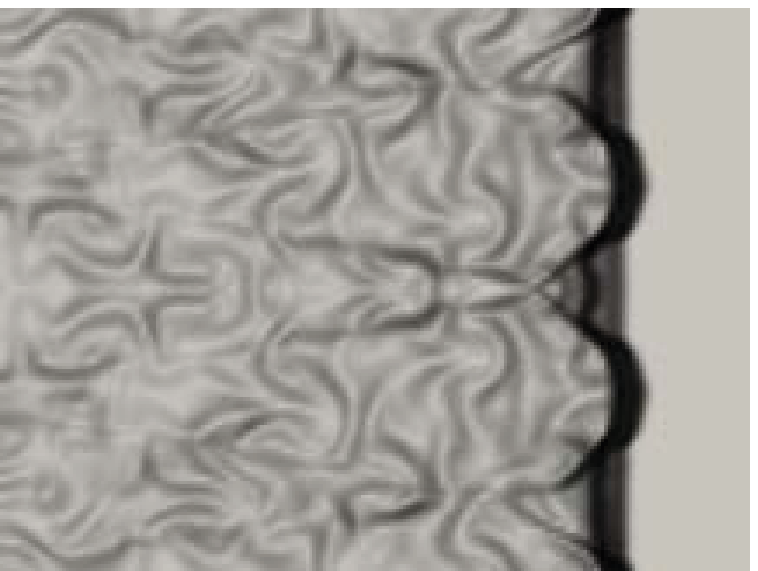}
        \includegraphics[height=2cm]{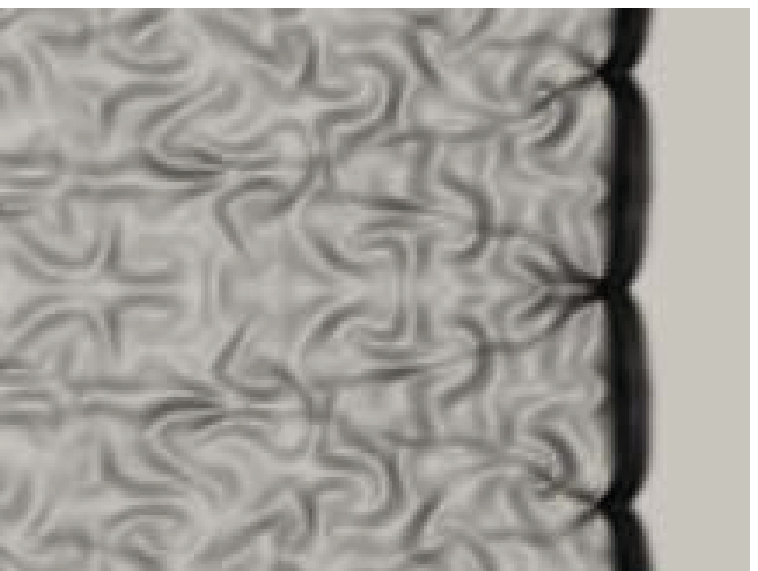}\\
	\caption{Evolution of the pressure and temperature in a 2D detonation simulation}
	\label{f:det2d}
\end{center}
\end{figure}

The evolution of the pressure and temperature of the two-dimensional flow field is shown in Figure~\ref{f:det2d}. The flow is modeled using 9 species gas mixture with 38 reactions. The mechanism used for the simulation is shown in Appendix A. The computational domain is rectangular with a length of 7.5 cm. The grid spacing in both directions is 50 $\mu$m. The detonation cells, between the shock and the multiple triple points in transverse motion, is clearly seen. This well-known cellular structure has been observed both in experiments and numerical simulations. We will show in the next section how the superior performance of the GPU can enhance our ability in modeling reactive flows.

\subsection{Performance Results}
\begin{figure}
\begin{center}
	\input{speedup1.tex}
	\includegraphics{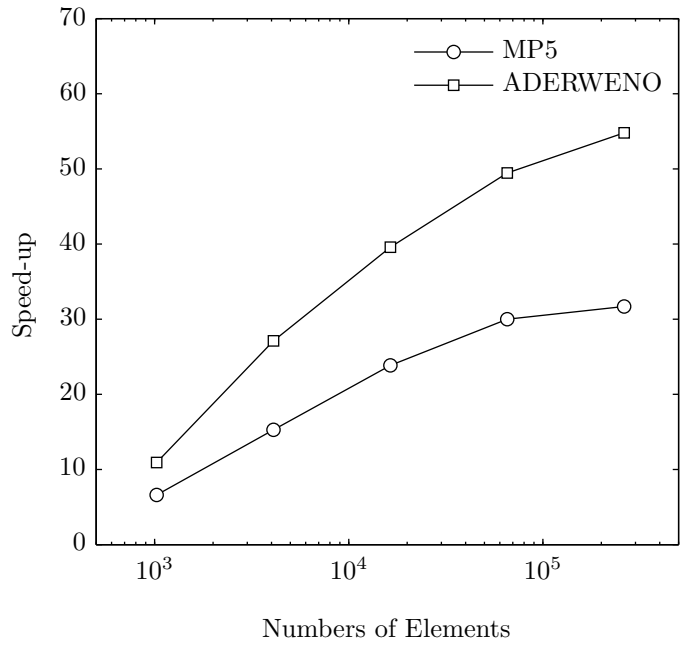}
	\caption{Performance of the fluid dynamics simulation only}
	\label{f:ideal_gas}
\end{center}
\end{figure}

Figure~\ref{f:ideal_gas} shows the performance of the solver for the simulation considering only the fluid dynamics aspect using the MP5
and the ADERWENO schemes. The speed-ups obtained in both cases are very promising. Since the ADERWENO scheme only requires single-stage time integration, it is faster than the MP5 scheme. For the ADERWENO scheme, we can obtain almost 60 times speed-up for a large grid which is about twice faster than the MP5 scheme. All the comparisons are made between a Tesla C2070 and an Intel Xeon X5650 (single thread) using double precision calculation. 

\begin{figure}
\begin{center}
	\input{memory_bandwidth.tex}
       \includegraphics[scale=1]{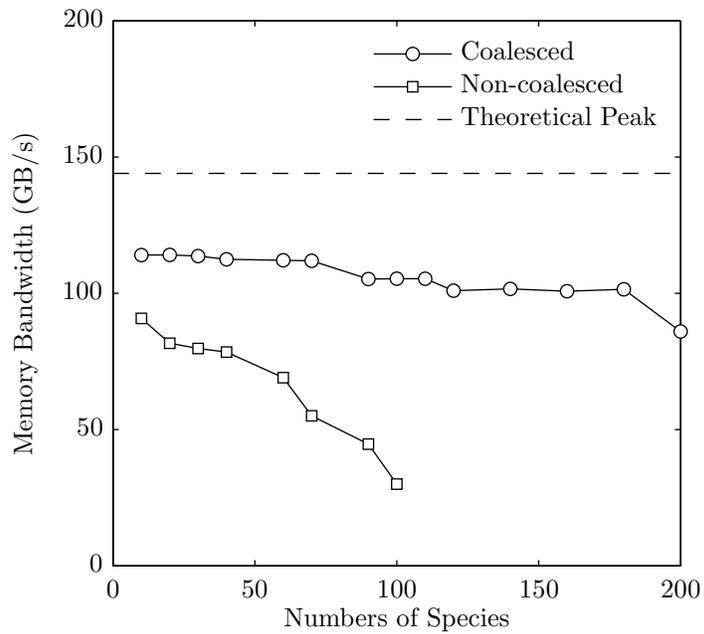}
	\captionof{figure}{Memory bandwidth of the kinetics solver measured on a Tesla C2050}
	\label{f:memory_bandwidth}
\end{center}
\end{figure}

The performance of the kinetics solver depends strongly on the memory access efficiency. Since the Gaussian elimination algorithm requires issuing a large amount of memory instructions (both \textit{Read} and \textit{Write}) to modify all the entries of the Jacobian, it is important to achieve high memory bandwidth while maintaining sufficient independent arithmetic operations to hide memory latency. This factor is very crucial in the case when the whole Jacobian is stored inside global memory (DRAM) since the DRAM latency is much higher than shared memory. Figure~\ref{f:memory_bandwidth} illustrates the memory access efficiency of the GPU kernel performing the Gaussian elimination procedure. It is clearly shown that coalesced memory access results in much higher memory bandwidth comparing to the non-coalesced pattern for the same number of operations. The memory bandwidth recorded for the kernel is up to 80\% of the theoretical peak limit of the device (144 GB/s for a Tesla C2050). The test is performance for a number of species up to 200. The result shown in figure~\ref{f:memory_bandwidth} indicates that the global memory access pattern implemented in the first approach is very efficient. In the second approach, we utilized shared memory to compensate for DRAM latency. However, due to the memory intensive nature of the chemical kinetics problem and the limitation of shared memory storage, there is a substantial amount of DRAM access which cannot be avoided. The memory bottleneck introduces addition memory latency which can affect the performance of the kernel.

\begin{figure}
\begin{center}
	\input{Implicit_solver_global_vs_shared.tex}
        \includegraphics[scale=.68]{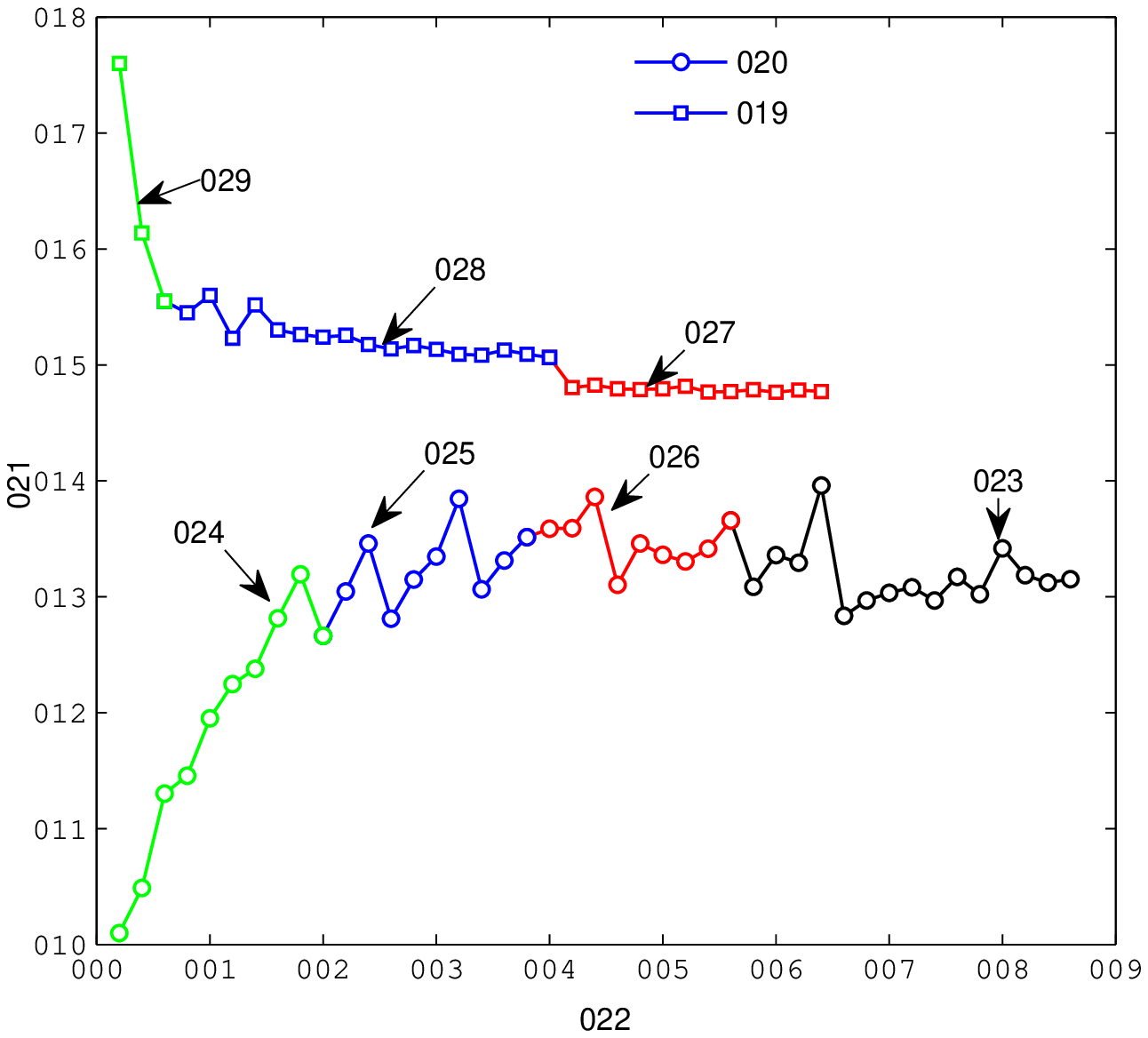}
	\captionof{figure}{Comparison of the speed-up factor obtained from the kinetics solver using both global and shared memory}
	\label{f:Implicit_solver_global_vs_shared}
\end{center}
\end{figure}

Figure~\ref{f:Implicit_solver_global_vs_shared} shows the performance of the kinetics solver only (i.e., no convection). 
Although the construction of the Jacobian and chemical source terms can also be a very time consuming process, we are interested here in the most computationally intensive part, i.e. the Gaussian elimination and its scaling. 
Hence, the performance is measured by solving a number of linear systems $A\cdot x = b$ with different system sizes ($N_s$) and grid sizes ($Ncell$).

We described two different implementation approaches in our earlier discussion. The first approach is to store everything on global memory and try to achieve high memory bandwidth by coalesced memory reads. The second approach is to transfer memory to shared memory for each row elimination. Figure~\ref{f:Implicit_solver_global_vs_shared} clearly shows that the global version outperforms the shared memory version in all cases. Since the shared memory approach requires additional memory transfers between each row elimination, it is only effective when $N_s$ is large. It is shown in the plot that the algorithm is only effective when $N_s > 100$. In contrast, the performance of the global memory version depends strongly on the size of the grid. Since the parallelization is achieved across all the computational cells, solving a large number of systems makes it much more efficient. The global memory version seems to perform well in all cases. The speed-up obtained is at least 30.

\begin{figure}
\begin{center}
	\input{chem_stencil.tex}
        \includegraphics[scale=1]{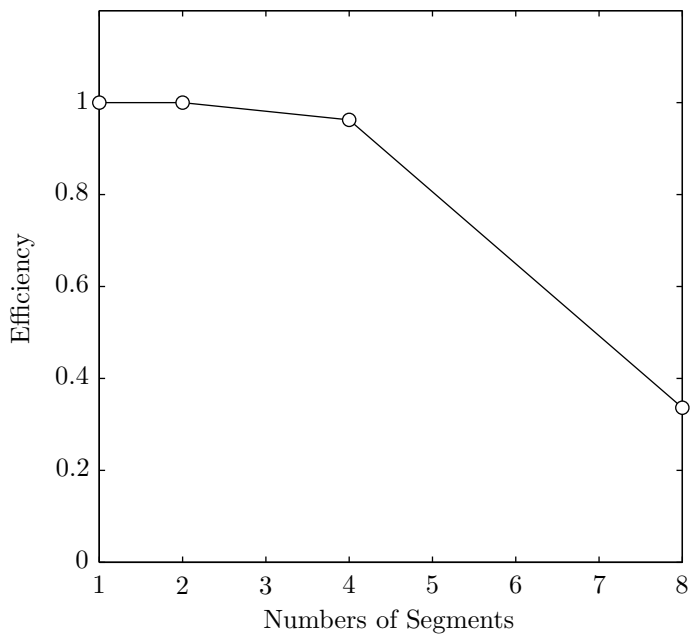}
	\captionof{figure}{Kinetics calculation of a $128 \times 128$ grid with 100 species. The calculation is done by breaking the computational domain into numbers of segments.}
	\label{f:chem_stencil}
\end{center}
\end{figure}

We were forced in this study to vary the grid size as the number of species increases, due to memory constraints. Thus, we were able to consider a $512 \times 512$ grid of cells for up to 25 species, then a $128 \times 128$ grid was possible up to 200 species, etc. The intense memory usage comes from the need to store a $N_s\times N_s$ Jacobian for each computational cell (e.g. this variable alone adds-up to more than 5 GB of global memory for 200 species on the $128 \times 128$ grid). This can be alleviated by storing only the Jacobian for a segment of the domain; for example, the kernel for a $512 \times 128$ grid can be called 4 times to solve the kinetics of a $512 \times 512$ grid. The constraint for this approach is that the segment itself must be sufficiently large, i.e. of the other of the number of streaming processors in the GPU, and that the overhead for each kernel call is small. This is shown in Figure~\ref{f:chem_stencil} where the calculation is performed for a grid with a fixed size of $128 \times 128$ with 100 species and varying number of segments. In this test case, the difference in execution time is negligible up to 4 segments. When the domain is further divided in 8 segments, the size of each segment gets smaller (2048 cells) and this approach is no longer effective.

\begin{figure}
\begin{center}
	\input{reactive_9sps.tex}
        \includegraphics[scale=1]{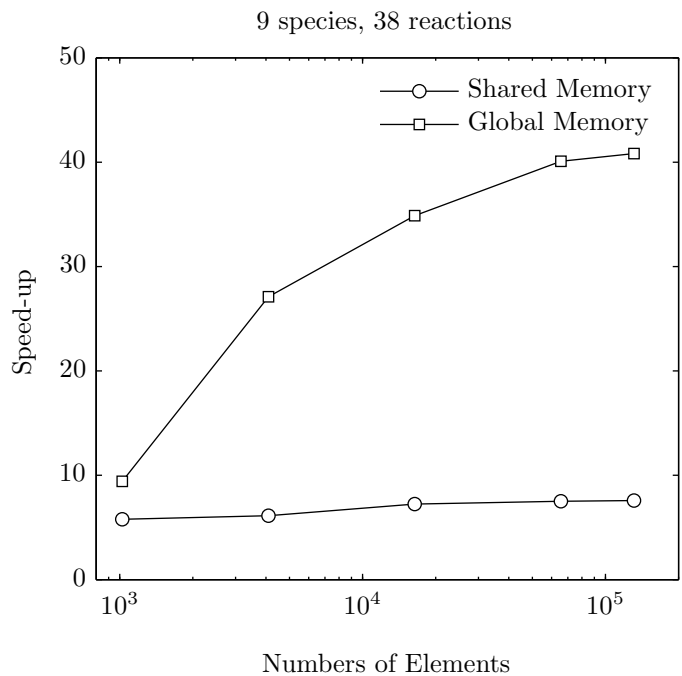}
	\captionof{figure}{Performance of the reactive flow solver for a 9-species gas}
	\label{f:reactive_9sps}
\end{center}
\end{figure}

Figure~\ref{f:reactive_9sps} shows the performance of the flow solver coupling with the chemical kinetics. The result shown in the plot is for a 9-species gas (hydrogen-air) with a mechanism consisting of 38 reactions (both forward and backward). MP5 scheme is utilized in all cases. Although ADERWENO scheme has shown to be faster than MP5, we do not expect to see significant difference in the performance because the computation time is dominated by chemical kinetics. The overall performance should be greatly dependent on the performance of the kinetics solver. It is clear that the global memory version outperforms the shared memory version. This is consistent with the results obtained earlier for the fluid dynamics and the kinetics separately. As the the grid size increases, the speed-up factor obtained for the shared memory version of the kinetics solver does not change rapidly. In contrast, the global memory version results in a 40 times speed-up for a large grid. 

\begin{figure}
\begin{center}
	\input{reactive_9vs36sps.tex}
        \includegraphics[scale=1]{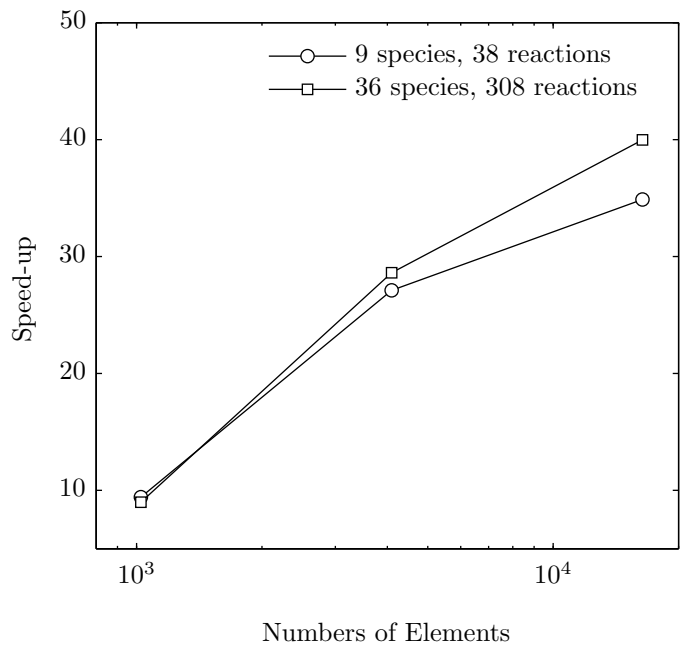}
	\captionof{figure}{Performance of the reactive flow solver for two different chemistry mechanism}
	\label{f:reactive_9vs36sps}
\end{center}
\end{figure}

We extend the simulation to model a larger mechanism with 36 species and 308 reactions (reduced hydrocarbon). The performance is compared with previous result of hydrogen-air. As illustrated in Figure~\ref{f:reactive_9vs36sps}, the result for the reduced hydrocarbon mechanism is faster than the hydrogen-air, presumably due to better performance in the construction of the chemical kinetics Jacobian.

\section{Conclusion and Future Works}
In the current paper, we described the implementation of a numerical solver for simulating chemically 
reacting flow on the GPU. The fluid dynamics is modeled using high-order shock-capturing schemes,
and the chemical kinetics is solved using an implicit solver. Results of both the fluid dynamics and chemical kinetics are shown. Considering only the fluid dynamics, we obtained a speed-up of 30 and 55 times compared to the CPU version for the MP5 and ADERWENO scheme, respectively. For the chemical kinetics, we presented two different approaches on implementing the Gaussian elimination algorithm on the GPU. The best performance obtained by solving the kinetics problem ranges from 30-40 depending on the size of the reaction mechanism. When the fluid dynamics is coupled with the kinetics, we obtained a speed-up factor of 40 times for a 9-species gas mixture with 38 reactions. The solver is also tested with a larger mechanism (36 species, 308 reactions) and the performance obtained is faster than the small mechanism. 

The current work can be extended in different ways. First, since the framework is performing well in shared memory architecture, it is possible to also extend it to distributed memory architecture utilizing Message Passing Interface (MPI). The extension permits using multi-GPU which is attracted for performing large-scale simulations. On the other hand, although the current simulation is done for chemically reacting flow, it is desired to extend it to simulate ionized gas (i.e. plasma) which requires modeling additional physical process (Collisional-Radiative) to characterize different excitation levels of the charged species. In adidition, the governing equations also need to be extended to characterize the thermal non-equilibrium environment of the plasma. Given that the physics has been well established\cite{kapper:ar_cr1,kapper:ar_cr2,JLC_plasma_molecularCR}, the extension is certainly achievable.

\pagebreak
\bibliographystyle{acm}	
\bibliography{bibtex_database}	

\begin{thebibliography}{10}

\bibitem{GPU_Brandvik}
{\sc Brandvik, T., and Pullan, G.}
\newblock {Acceleration of a 3D Euler Solver using Commodity Graphics
  Hardware}.
\newblock In {\em 46th AIAA Aerospace Sciences Meeting\/} (2008), AIAA paper
  08-607.

\bibitem{JLC_supersonic}
{\sc Cambier, J.-L., Adelman, H.~G., and Menees, G.~P.}
\newblock {Numerical Simulations of Oblique Detonations in Supersonic
  Combustion Chamber}.
\newblock {\em J. of Prop. and Power 5}, 4 (1989), 481--491.

\bibitem{JLC_plasma_molecularCR}
{\sc Cambier, J.-L., and Moreau, S.}
\newblock {Simulations of a Molecular Plasma in Collisional-Radiative
  Nonequilibrium}.
\newblock AIAA paper 93-3196.

\bibitem{Einfeldt:HLLE}
{\sc Einfeldt, B., Munz, C.~D., Roe, P.~L., and Sj\"{o}green, B.}
\newblock {On Godunov-Type Methods Near Low Densities}.
\newblock {\em J. Comp. Phys. 92}, 2 (1991), 273--295.

\bibitem{GPU_NSSUS}
{\sc Elsen, E., LeGresley, P., and Darve, E.}
\newblock {Large Calculation of the Flow Over a Hypersonic Vehicle Using a
  GPU}.
\newblock {\em J. Comp. Phys. 227}, 24 (2008), 10148--10161.

\bibitem{RT}
{\sc Gardner, C.~L., Glimm, J., McBryan, O., Menikoff, R., Sharp, D.~H., and
  Zhang, Q.}
\newblock {The Dynamics of Bubble Growth for Rayleigh-Taylor Unstable
  Interfaces}.
\newblock {\em Physics of Fluid 31}, 3 (1988), 447--465.

\bibitem{kapper:ar_cr1}
{\sc Kapper, M.~G., and Cambier, J.-L.}
\newblock {Ionizing Shocks in Argon. Part I: Collisional-Radiative Model and
  Steady-State Structure}.
\newblock {\em Journal of Applied Physics 109}, 11 (2011), 113308.

\bibitem{kapper:ar_cr2}
{\sc Kapper, M.~G., and Cambier, J.-L.}
\newblock {Ionizing Shocks in Argon. Part II: Transient and Multi-Dimensional
  Effects}.
\newblock {\em Journal of Applied Physics 109}, 11 (2011), 113309.

\bibitem{Klockner20097863}
{\sc Klockner, A., Warburton, T., Bridge, J., and Hesthaven, J.}
\newblock {Nodal Discontinuous Galerkin Methods on Graphics Processors}.
\newblock {\em J. Comp. Phys 228}, 21 (2009), 7863 -- 7882.

\bibitem{CUDA_programming_guide}
{\sc {NVIDIA Corporation}}.
\newblock {\em {Compute Unified Device Architecture Programming Guide version
  4.0}}.
\newblock 2011.

\bibitem{suresh:mp5}
{\sc Suresh, A., and Huynh, H.~T.}
\newblock Accurate monotonicity-preserving schemes with runge-kutta time
  stepping.
\newblock {\em J. Comp. Phys. 136\/} (1997), 83--99.

\bibitem{titarev:ader}
{\sc Titarev, V.~A., and Toro, E.~F.}
\newblock {ADER} schemes for three-dimensional nonlinear hyperbolic systems.
\newblock {\em J. Comp. Phys 204}, 2 (2005), 715--736.

\bibitem{toro:Riemann_solver}
{\sc Toro, E.~F.}
\newblock {\em Riemann solvers and numerical methods for fluid dynamics - {A}
  practical introduction - 2nd edition}.
\newblock Springer, 1999.

\bibitem{Vandyke}
{\sc Van~Dyke, M.}
\newblock {\em An Album of Fluid Motion}.
\newblock Parabolic Press, Inc., 1989.

\bibitem{Volkov}
{\sc Volkov, V.}
\newblock {Better Performance at Lower Occupancy}.
\newblock In {\em GPU Technology Conference\/} (2010).

\bibitem{Wong20112132}
{\sc Wong, H.-C., Wong, U.-H., Feng, X., and Tang, Z.}
\newblock {Efficient Magnetohydrodynamic Simulations on Graphics Processing
  Units with CUDA}.
\newblock {\em Computer Physics Communications 182}, 10 (2011), 2132 -- 2160.

\bibitem{Woodward1984115}
{\sc Woodward, P., and Colella, P.}
\newblock The numerical simulation of two-dimensional fluid flow with strong
  shocks.
\newblock {\em J. Comp. Phys 54}, 1 (1984), 115--173.

\end{thebibliography}

\pagebreak
\section*{Appendix A}
\begin{center}Reaction Mechanism for Hydrogen-Air Detonation\end{center}
\begin{center}
	\begin{tabular}{|r|c|}
\hline
 1   &    $H + O_2        \leftrightarrow O + OH$\\ \hline
 2   &    $O + H_2        \leftrightarrow H + OH$\\ \hline
 3   &    $H_2 + OH       \leftrightarrow H + H_2O$\\ \hline
 4   &    $OH + OH       \leftrightarrow H_2O + O$\\ \hline
 5   &    $H + OH   + M  \leftrightarrow H_2O     + M$\\ \hline
 6   &    $H + H    + M  \leftrightarrow H_2      + M$\\ \hline
 7   &    $H + O    + M  \leftrightarrow OH      + M$\\ \hline
 8   &    $2O    + M  	 \leftrightarrow O_2      + M$\\ \hline
 9   &    $H_2 + O_2       \leftrightarrow HO_2 + H$\\ \hline
10   &    $H + O_2   + M  \leftrightarrow HO_2     + M$\\ \hline
11   &    $H + HO_2       \leftrightarrow OH + OH$\\ \hline
12   &    $H + HO_2       \leftrightarrow O + H_2O$\\ \hline 
13   &    $O + HO_2       \leftrightarrow O_2 + OH$\\ \hline
14   &    $OH + HO_2      \leftrightarrow O_2 + H_2O$\\ \hline
15   &    $H_2O_2     + M  \leftrightarrow OH + OH + M$\\ \hline
16   &    $HO_2 + HO_2     \leftrightarrow H_2O_2 + O_2$\\ \hline
17   &    $H + H_2O_2      \leftrightarrow H_2 + HO_2$\\ \hline
18   &    $O + H_2O_2      \leftrightarrow OH + HO_2$\\ \hline
19   &    $OH + H_2O_2     \leftrightarrow H_2O + HO_2$\\ \hline
	\end{tabular}
	\label{tab:}
\end{center}

%\pagebreak

\end{document}